\documentclass[12pt]{iopart}

\usepackage{iopams} 
\usepackage{bm,graphics,graphicx,epsfig,color}

\begin{document}

\title[Light Deflection]{The 1919 measurement of the deflection of light}

\author{Clifford M. Will$^{1,2}$}

\ead{cmw@physics.ufl.edu}
\address{
$^1$ Department of Physics,
University of Florida, Gainesville FL 32611, USA
 \\
$^2$ GReCO, Institut d'Astrophysique de Paris, UMR 7095-CNRS,
Universit\'e Pierre et Marie Curie, 98$^{bis}$ Bd. Arago, 75014 Paris, France
}

\begin{abstract}
The measurement of the deflection of starlight during a total solar eclipse on May 29, 1919 was the first verification of general relativity by an external team of scientists, brought Einstein and his theory to the attention of the general public, and left a legacy of experimental testing that continues today.  The discovery of gravitational lenses turned Einstein's deflection into an important tool for astronomy and cosmology.  This article reviews the history of the 1919 measurement and other eclipse measurements, describes modern measurements of the effect using radio astronomy, and of its cousin, the Shapiro time delay, and discusses gravitational lenses.
\end{abstract}


\maketitle

\section{Introduction}
\label{sec:intro}
The headline in the London Times of November 7, 1919 read ``Revolution in Science/ New Theory of the Universe/ Newtonian Ideas Overthrown.'' It signaled a brave new world in which the old values of absolute space and absolute time were lost forever. To some emerging from the devastation of the First World War, it meant the overthrow of all absolute standards, whether in morality or philosophy, music or art. In a 1983 survey of twentieth-century history, the British historian Paul Johnson argued that the ``modern era'' began not in 1900, not in August 1914, but with the event that spawned this headline~\cite{pauljohnson}.

The event that caused such an uproar was the successful measurement of the bending of starlight by the Sun during a solar eclipse in May 1919~\cite{1920RSPTA.220..291D}.  The amount of bending agreed with the prediction of Einstein's general theory of relativity, but disagreed with the prediction of Newton's gravitational theory.   The announcement made Einstein an overnight international celebrity and brought the strange concepts and complex mathematics of general relativity before the general public.   

It also brought to the fore the centrality of experiment in verifying or falsifying a fundamental theory of nature.  It set off a flurry of activity worldwide to test the so-called three critical predictions of general relativity: the deflection of light, the gravitational redshift, and the perihelion advance of Mercury.    But during the next 20 years, the results were not promising.  Subsequent measurements of the deflection of light ranged between three-quarters and one and one-half times the general relativistic prediction, and the precisions were low.  The gravitational redshift was never measured reliably.  And while the predicted relativistic advance of Mercury's perihelion appeared to resolve a problem dating back to the 1850s, some skeptical astronomers continued to hunt for Vulcan, a hypothetical planet between Mercury and the Sun, whose gravitational perturbations would solve the problem without relativity (see~\cite{Crelinsten} for a detailed history of this period). 

This failure to ``seal the deal'' empirically with general relativity was partially responsible for the steady decline of scientific interest in the theory from the 1920s onward.   It wasn't until the 1960s that a renaissance in experimental relativity occurred, spurred by astronomical discoveries that heralded a central role for the theory in astrophysics, by advances in precision technology that would provide tools to measure the minuscule effects predicted by the theory, and by the space program, which would provide a new platform for many tests.

The deflection of light, and its related effect, the Shapiro time delay, played a key role in this renaissance, and today these effects are measured with precisions of hundredths to thousandths of a percent, consistently in agreement with general relativity, and may soon reach the part-per-million level.   And as we celebrate the centenary of general relativity, Einstein's deflection is still front and center.  Morphed into the gravitational lens, it is 
a prime tool in efforts to map dark matter, study galactic structure, find exoplanets, probe dark energy and trace fluctuations in the cosmic background radiation.  

In this article, we shall review this first of our selected ``milestones of general relativity''.  We begin in Sec.~\ref{sec:eclipse} by describing the 1919 measurement itself, both its background and its immediate aftermath.  In Sec.~\ref{sec:struggles} we discuss the years following the measurement, characterized by skepticism of general relativity in some circles and by not terribly conclusive experimental results.  Section \ref{sec:modern} reviews the modern high-precision measurements, and Sec.~\ref{sec:legacy} describes the broad legacy of 1919 in experimental gravity.  In Sec.~\ref{sec:lenses} we describe gravitational lenses, while Sec.~\ref{sec:conclusions} makes concluding remarks.

\section{The 1919 eclipse expedition: Background and aftermath}
\label{sec:eclipse}

From the moment that Einstein recognized the equivalence between gravity and accelerated frames, he realized that gravity would affect the trajectory of light.  In 1911, he determined that the deflection of a light ray grazing the Sun should be 0.875 arcseconds~\cite{1911AnP...340..898E}.  He proposed that the effect be looked for during a total solar eclipse, during which stars near the Sun would be visible and any bending of their rays could be detected through the displacement of the stars from their normal positions.   Several teams, including one headed by Erwin Finlay-Freundlich of the Berlin Observatory and one headed by William Campbell of the Lick Observatory in the USA headed for the Crimea to observe the eclipse of August 21, 1914.  But World War I intervened, and Russia sent many of the astronomers home, interned others, and temporarily confiscated much of the equipment; the weather was too bad to permit useful observations anyway.  

In November of 1915, armed with the full theory of general relativity, Einstein realized that the predicted deflection was double his earlier value.   From a modern perspective we understand that his original 1911 derivation was correct as it stood.  It is equivalent to the deflection that would be calculated using Newtonian theory for a particle moving at the speed of light.  This ``Newtonian deflection'' was first calculated by Henry Cavendish around 1784, and by Johann von Soldner in 1803~\cite{willcavendish}.   The doubling comes about because the Newtonian deflection must be augmented by the bending of locally straight lines relative to straight lines very far from the Sun, a result of spatial curvature. The fact that the two effects have the same size is a feature of general relativity; in alternative theories of gravity, the Newtonian effect is the same, but the space curvature effect may differ from theory to theory.
  
This doubling meant that the effect was now more accessible to observation. But the fact that a successful observation came as early as 1919, only three years after publication of the complete general theory, must be credited to Arthur Stanley Eddington.  By the time of the outbreak of World War I, Eddington was the Plumian Professor at Cambridge University, and one of the foremost observational astronomers of the day.  The war had effectively stopped direct communication between British and German scientists, but the Dutch cosmologist Willem de Sitter managed to forward to Eddington Einstein's latest paper together with several of his own on the general theory of relativity.  In 1917, Eddington prepared a detailed and laudatory report on the theory for the Physical Society of London, and began to work with Astronomer Royal Frank Dyson on an eclipse expedition to measure the predicted deflection of light.  Dyson had pointed out that the eclipse of May 29, 1919 would be an excellent opportunity because of the large number of bright stars expected to form the field around the Sun.   A grant of 1,000 pounds sterling was obtained from the government, and planning began in earnest.  The outcome of the war was still in doubt at this time, and a danger arose that Eddington would be drafted. As a devout Quaker, he had pleaded exemption from military service as a conscientious objector, but, in its desperate need for more manpower, the Ministry of National Service appealed the exemption. Finally, after three hearings and a last-minute appeal from Dyson, the exemption from service was upheld on July 11, 1918.  This was just one week before the second Battle of the Marne, a turning point in the war.
The decision reeks of irony: a British government permitting a pacifist scientist to avoid wartime military duty so that he could go off and try to verify a theory produced by an enemy scientist.

On March 8, 1919, just four months after the armistice ending the war, two expeditions set sail from England: Eddington's for the island of Principe, off the coast of present-day Equatorial Guinea; the other team under Andrew Crommelin for the city of Sobral, in northern Brazil.  The principle of the experiment is deceptively simple. During a total solar eclipse, the moon hides the Sun completely, revealing the field of stars around it. Using a telescope and photographic plates, the astronomers take pictures of the obscured Sun and the surrounding star field. These pictures are then compared with pictures of the same star field taken when the Sun is not present. The comparison pictures are taken at night, several months earlier or later than the eclipse, when the Sun is nowhere near that part of the sky and the stars are in their true, undeflected positions. In the eclipse pictures, the stars whose light is deflected would appear to be displaced away from the Sun relative to their actual positions.  Because the deflection varies inversely as the angular distance of the star from the Sun, stars far from the Sun establish the fixed reference points for comparing the sets of plates.

Because of turbulence in the Earth's atmosphere (``seeing''), starlight passing through it can suffer deflections comparable to the effect being measured.   But because they are random in nature, they can be averaged away if one has many images, revealing the underlying systematic displacements away from the Sun.   To this end, of course, it helps to have a clear sky.
But on the day of the eclipse at Eddington's site, a rainstorm started, and as the morning wore on, he began to lose all hope.  But at the last moment, the weather began to change for the better, and when the partial eclipse was well advanced, the astronomers began to get a glimpse of the Sun.
Of the sixteen photographs taken through the remaining cloud cover, only two had reliable images, totaling only about five stars.  Nevertheless, comparison of the two eclipse plates with a comparison plate taken at the Oxford University telescope before the expedition yielded results in agreement with general relativity, corresponding to a deflection at the limb of the Sun (grazing ray) of $1.60 \pm 0.31$ arcseconds, or $0.91 \pm 0.18$ times the Einsteinian prediction.  The Sobral expedition, blessed with better weather, managed to obtain eight usable plates showing at least seven stars each. The nineteen plates taken on a second telescope turned out to be worthless because the telescope apparently changed its focal length just before totality of the eclipse, possibly as a result of heating by the Sun. Analysis of the good plates yielded a grazing deflection of $1.98 \pm 0.12$ arcseconds, or $1.13 \pm 0.07$ times the predicted value~\cite{1920RSPTA.220..291D}.  

Before this, Einstein had been an obscure Swiss/German scientist, well known and respected within the small European community of physicists, but largely unknown to the outside world.  With the announcement of the measurement of the deflection at the Royal Society of London on November 6, 1919, all this changed, and Einstein and his theory became overnight sensations.   The Einstein aura has not abated since.

On the other hand, Einstein's fame did engender a backlash, especially in Germany.   In 1920, Paul Weyland organized a public forum in which Einstein and his theories were denounced, and Nobel Laureate Philipp Lenard had Soldner's 1803 article reprinted in an attempt to demonstrate that Einstein had committed plagiarism of an Aryan scientist's work.  Behind many of these attacks was antisemitism, and relativity was frequently characterized as ``Jewish science''.   The vast majority of non-Jewish German physicists did not share this view, however, and despite the Nazi takeover in Germany and the subsequent dismissal and emigration of many Jewish physicists (including Einstein), the anti-relativity program became little more than a footnote in the history of science.  

Other questions were raised, however, about the results of Eddington's measurements.  Given the poor quality of the data, did they really support Einstein or not?   Was it proper for Eddington to discard the data from the second telescope at the Sobral site?   Given Eddington's enthusiasm for the theory of general relativity, some wondered whether he had selected or massaged the data to get the desired result.   Numerous reanalyses between 1923 and 1956 of the plates used by Eddington yielded the same results within ten percent.   In 1979, on the occasion of the centenary of Einstein's birth, astronomers at the Royal Greenwich Observatory reanalysed both sets of Sobral plates using a modern tool called the Zeiss Ascorecord and its data reduction software~\cite{1979Obs....99..195H}.  The plates from the first telescope yielded virtually the same deflection as that obtained by Eddington's team with the errors reduced by 40 percent.   Despite the scale changes in the second set of Sobral plates, the analysis still gave a result $1.55 \pm 0.34$ arcseconds at the limb, consistent with general relativity, albeit with much larger errors.   Looking back on Eddington's treatment of the data, Kennefick~\cite{2009PhT....62c..37K} has argued that there is no credible evidence of bias on his part.

\section{Testing general relativity: the early struggles}
\label{sec:struggles}

The publicity surrounding Eddington's famous announcement has left the impression that his was the only test of the deflection using eclipse measurements, successful or otherwise.   But the history is much richer~\cite{Crelinsten}.  Campbell and Heber Curtis of the Lick Observatory analyzed plates from a 1900 eclipse in Georgia and a 1918 eclipse in Washington State in the USA and found no deflection; ironically they reported this negative result at the Royal Society of London meeting in July 1919 in the midst of Eddington's data analysis (at the meeting, rumors were already going around that Eddington would report a positive result).   Following up on Eddington's success, seven teams tried the measurement during a 1922 eclipse in Australia, although only three succeeded.   Campbell and Robert Trumpler of the Lick team reported a result for the deflection at the limb of $1.72 \pm 0.11$ arcseconds, while a Canadian team and an England/Australian team reported values between $1.2$ and $2.3$ arcseconds.  
Later eclipse measurements continued to support general relativity: one in 1929, two in 1936, one each in 1947 and 1952, and one in 1973.  Surprisingly, there was very little improvement in accuracy, with different measurements giving values anywhere between three-quarters and one and one-third times the general relativistic prediction, yet there was little doubt about Einstein beating Newton (for reviews see~\cite{1960VA......3...47V,BertottiBrillKrotkov}). 

The 1973 expedition is a case in point.  Organized by the University of Texas and Princeton University, the observation took place in June at Chinguetti Oasis in Mauritania\footnote{The author vividly remembers Bryce DeWitt's slide show on the expedition presented at the Les Houches Summer School on black holes in August 1974, including photos of DeWitt and Richard Matzner perched atop camels}. The observers had the benefit of 1970s technology: modern photographic emulsions, a temperature-controlled telescope shed, sophisticated motor drives to control the direction of the telescope accurately, and computerized analysis of the photographs. Unfortunately, they couldn't control the weather any better than Eddington. Eclipse morning brought high winds, drifting sand, and dust too thick to see the Sun. But as totality of the eclipse approached, the winds died down, the dust began to settle, and the astronomers took a sequence of photographs during what they have described as the shortest six minutes of their lives. They had hoped to gather over $1000$ star images, but the dust cut the visibility to less than $20$ percent and only a disappointing $150$ were obtained. After a follow-up expedition to the site in November to take comparison plates, the photographs were analyzed using the GALAXY Measuring Engine at the Greenwich Observatory, with a result $0.95 \pm 0.11$ times the general relativity prediction, essentially no improvement in accuracy over previous eclipse measurements~\cite{1976AJ.....81..452B,1976AJ.....81..455J}.

Although the deflection of light was gradually accepted as a success for general relativity, the second of Einstein's empirical pillars of the theory -- the gravitational redshift -- was not conclusively measured until the 1960s.   In 1917, Charles E. St. John of the Mount Wilson Observatory in California reported no relativistic shift of spectral lines from the Sun\footnote{This result apparently had a direct negative impact on Einstein's candidacy for the Nobel Prize that year. The prize would not be awarded to him until 1922, and then only for the photoelectric effect, not for any of his relativistic theories.}, and a 1918 report from an observatory in Kodiakanal in India was inconclusive.  This established a pattern that would last for decades, and that
was seized upon by some as reason to doubt the theory.   For example, the mathematician Alfred North Whitehead produced an alternative theory of gravity in 1922~\cite{whitehead22} designed to retain the flat space-time of special relativity while providing an ``action-at-a-distance'' tensor potential that would give the correct deflection of light and orbital motion of particles, while predicting no gravitational redshift effect.  

Unfortunately, the measurement of the solar shift  is not simple.
Solar spectral lines are subject to the ``limb effect'', a variation
of spectral line wavelengths between the center of the solar disk and
its edge or ``limb''; this effect is actually a Doppler shift caused by complex
convective and turbulent motions in the photosphere and lower
chromosphere, and is expected to be minimized by observing at the
solar limb, where the motions are predominantly transverse.  Pressure shifts are also important for certain elements.   The
secret is to use strong, symmetrical lines, leading to unambiguous
wavelength measurements.  Truly reliable measurements were not made until
1962 and 1972.  Finally, in 1991, LoPresto et
al.~\cite{lopresto91} 
measured the solar shift in agreement with the prediction to about 2 percent by 
observing the oxygen triplet lines both in
absorption in the limb and in emission just off the limb.  

Of course, the first true measurement of the gravitational redshift was the classic Pound-Rebka experiment of 1960~\cite{1960PhRvL...4..337P}, which
measured the frequency shift of gamma-ray photons from ${}^{57} \mathrm{Fe}$
as they ascended or descended the Jefferson Physical Laboratory
tower at Harvard University.  

The third pillar of general relativity -- the perihelion advance of Mercury -- was an immediate success for Einstein, yet it did not seem to play as large a role in the early debates over the validity of general relativity as did the deflection.  Ironically, eclipse measurements played a small role in the Mercury story as well, because of the planet Vulcan.   This hypothetical planet, lying somewhere between Mercury and the Sun, was proposed by Urbain Jean Le Verrier and others in the 1860s to explain the anomalous advance of Mercury's perihelion at $43$ arcseconds per century (the modern value) that Le Verrier had uncovered.  Numerous ``sightings'' of Vulcan were made during the last decades of the 19th century.   Many early eclipse expeditions, notably in 1901, 1905 and 1908, were partly motivated by the search for the planet Vulcan.  Einstein's explanation of the effect appeared to lay the issue to rest, although some diehard critics of general relativity continued to promote Vulcan, and some later eclipse expeditions continued to devote effort in search of Vulcan.   No credible evidence for the planet was ever found. 

Beginning in the 1960s a renaissance for general relativity and its experimental tests began, and the deflection of light, along with its cousin, the Shapiro time delay, played a central role.  In the next section, we turn to the modern story of the deflection of light.

\section{Modern measurements of the deflection}
\label{sec:modern}

Modern measurements of the deflection of light as well as other weak-field tests of gravity are cast in the language of the parametrized post-Newtonian (PPN) formalism, which parametrizes the weak-field, slow-motion metric of a class of alternative theories of gravity in terms of dimensionless parameters $\gamma$, $\beta$, $\xi$, $\alpha_1$, $\alpha_2$, etc., whose values vary from theory to theory.  In general relativity $\gamma =1$, $\beta =1$, while the other parameters all vanish.  It is $\gamma$ that is related to space curvature, whose effect must be added to the Newtonian deflection to get the full prediction.

\begin{figure}[t]
\centering
  \includegraphics[width=4in]{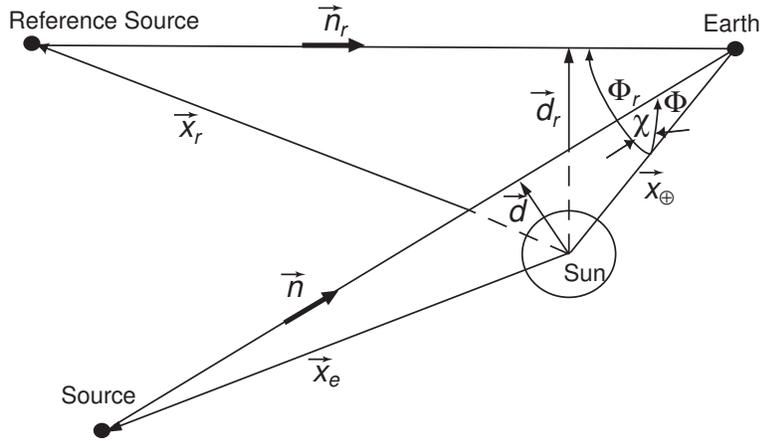}
  \caption{Geometry of light deflection measurements.}
  \label{deflectiongeom}
\end{figure}

In this framework, a light ray which passes the Sun at a distance $d$ is
deflected by an angle
\begin{equation}
  \delta \theta = \frac{1}{2} (1 + \gamma) \frac{4  GM_\odot}{c^2 d}
  \frac{1 + \cos \Phi}{2} \,,
  \label{E28}
\end{equation}
where $M_\odot$ is the mass of the Sun, $G$ is Newton's gravitational constant, $c$ is the speed of light
and $\Phi$ is the angle between the Earth-Sun line and the
incoming direction of the photon (see Figure~\ref{deflectiongeom}).
For a grazing ray, $\Phi \approx 0$,
$d \approx R_\odot$, where $R_\odot$ is the solar radius, and
\begin{equation}
  \delta \theta \approx \frac{1}{2} (1 + \gamma) 1.''7505,
  \label{E29}
\end{equation}

The development of radio interferometery, and later
of very-long-baseline radio interfer\-om\-etry (VLBI), produced
greatly improved determinations of
the deflection of light. These techniques now have the capability
of measuring angular separations and changes in angles
to accuracies better than 
100 microarcseconds. Early measurements took advantage of
the fact that
groups of strong quasars annually pass
very close to the Sun (as seen from the Earth), such as the
group 3C273, 3C279 and 3C48. As the Earth moves in its orbit, changing the
lines of sight of the quasars relative to the Sun, the angular
separation between pairs of quasars
varies. 
A number of measurements of this kind over the period 1969\,--\,1975 yielded
accurate determinations of the coefficient $\frac{1}{2}(1+ \gamma)$, or equivalently $\gamma -1$, reaching levels of a percent.
A 1995 VLBI measurement using 3C273 and
3C279 yielded $\gamma -1=(-8 \pm 34) \times 10^{-4}$~\cite{lebach}, while a 2009 measurement using the VLBA targeting the same two quasars plus two other nearby radio sources yielded 
$\gamma -1=(-2 \pm 3)  \times 10^{-4}$~\cite{2009ApJ...699.1395F}.

\begin{figure}[t]
\centering
 \includegraphics[width=4in]{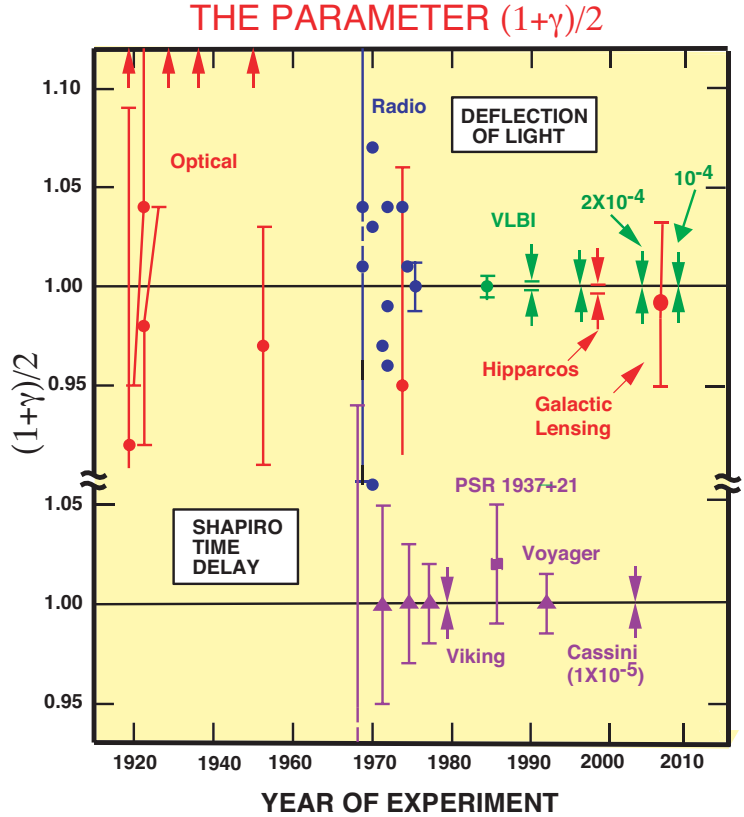}
  \caption{Measurements of the coefficient $(1 + \gamma )/2$ from
    light deflection and time delay measurements. Its GR
    value is unity. The arrows at the top denote anomalously large
    values from early eclipse expeditions. The Shapiro time-delay
    measurements using the Cassini spacecraft yielded an agreement with GR
    to $10^{-3}$~percent, and VLBI light deflection measurements have
    reached 0.01~percent. Hipparcos denotes the optical astrometry
    satellite, which reached 0.1~percent.}
  \label{gammavalues}
\end{figure}

In recent years, transcontinental and intercontinental VLBI observations 
of quasars and
radio galaxies have been 
made primarily to monitor the Earth's rotation
(``VLBI'' in Figure~\ref{gammavalues}). These measurements are
sensitive to the deflection of light over
almost the entire celestial sphere (at $90 ^\circ$ from the Sun, the
deflection is still 4 milli\-arcseconds).
A 2004 analysis of  almost 2 million VLBI observations
of 541 radio sources, made by 87 VLBI sites
yielded 
$\gamma-1= (-1.7 \pm 4.5) \times 10^{-4}$~\cite{sshapiro04}.
Analyses that incorporated data through 2010 yielded $\gamma-1= (-0.8 \pm 1.2) \times 10^{-4}$~\cite{2009A&A...499..331L,2011A&A...529A..70L}.  

To reach high precision at optical wavelengths requires observations from space.
The Hipparcos optical astrometry
satellite yielded a measurement of the deflection at the level of 0.3
percent~\cite{hipparcos}.  GAIA, a high-precision astrometric orbiting telescope launched by ESA in 2013~\cite{gaia} possesses astrometric capability ranging from 10 to a few hundred microarcseconds, plus the ability measure the locations of a billion stars down to 20th magnitude; it could
measure the light-deflection and
$\gamma$ to the $10^{-6}$ level~\cite{2002EAS.....2..107M}. 

Complementary to the deflection of light is the Shapiro time delay, an
additional non-Newtonian delay in the round-trip travel time of a signal sent, say, from the Earth to a distant planet or spacecraft and back,
given by (see Figure~\ref{deflectiongeom})
\begin{equation}
  \delta t = 2 (1 + \gamma) \frac{GM_\odot}{c^2}
  \ln \left( \frac{(r_\oplus + {\bf x}_\oplus \cdot {\bf n})
  (r_\mathrm{e} - {\bf x}_\mathrm{e} \cdot {\bf n})}{d^2} \right),
  \label{E31}
\end{equation}
where $ {\bf x}_\mathrm{e} $ ($ {\bf x}_\oplus $) are the vectors, and
$ r_\mathrm{e} $ ($ r_\oplus $) are the distances from the Sun
to the source (Earth), respectively. For a ray
which passes close to the Sun,
\begin{equation}
  \delta t \approx \frac{1}{2} (1 + \gamma)
  \left [ 240 - 20 \ln \left ( \frac{d^2}{r} \right ) \right ] \mathrm{\ \mu s},
  \label{E32}
\end{equation}
where $d$ is the distance of closest approach of the ray in solar
radii, and $r$ is the distance of the planet or satellite from the
Sun, in astronomical units.

This was not predicted by Einstein.  It was found in 1964 by radio astronomer Irwin Shapiro, who  calculated it while pondering the problem of bouncing radar signals from Venus and Mercury~\cite{1964PhRvL..13..789S}.  He then made the first measurement of the effect.
During the next four decades numerous high-precision measurements were made of the Shapiro delay
using radar ranging to targets passing behind the Sun  (superior
conjunction). 
The targets employed included
planets, such as Mercury or Venus, and
artificial satellites, such as Mariners~6 and 7, Voyager~2,
the Viking
Mars landers and orbiters, and the Cassini spacecraft to Saturn.

The results for the coefficient $\frac{1}{2}(1+ \gamma)$
of all radar time-delay measurements
performed to date (including a measurement of the one-way time delay
of signals from the millisecond pulsar PSR 1937+21)
are shown in Figure~\ref{gammavalues}. The 1976 Viking experiment resulted in a
0.1 percent measurement~\cite{reasenberg}. 
A significant improvement was reported in 2003 
from Doppler tracking of the Cassini
spacecraft while it was on its way to Saturn~\cite{bertotti03}, with a
result $\gamma -1 = (2.1 \pm 2.3) \times 10^{-5}$.  This was made possible
by the ability to do Doppler measurements using both X-band (7175~MHz) and
Ka-band (34316~MHz) radar, thereby significantly reducing the dispersive
effects of the
solar corona (in using Doppler tracking, one is essentially measuring the rate of change of the Shapiro delay). In addition, the 2002 superior conjunction of Cassini was
particularly favorable: with the spacecraft at 8.43 astronomical units from
the Sun, the distance of closest approach of the radar signals to the Sun
was only $1.6 \, R_\odot$.

\section{Putting GR to the test: A legacy of 1919}
\label{sec:legacy}

It is not the purpose of this article to give a full overview of experimental tests of general relativity; readers are referred elsewhere for such reviews~\cite{shapiro,2008ARNPS..58..207T,2010AmJPh..78.1240W,2014LRR....17....4W,Agashe:2014kda}.  Instead we shall give here a list of selected measurements or observations that illustrate the breadth and depth of tests of general relativity that form the legacy of 1919:

\begin{itemize}
\item
Global fits of solar system orbital data to verify the relativistic perihelion precession of Mercury to a few parts in $10^5$, coupled with helioseismology measurements showing that the solar quadrupole moment is too small to have an effect at this level~\cite{2014A&A...561A.115V}. 

\item
Verification of Einstein's gravitational redshift to parts in $10^4$ using a hydrogen maser atomic clock on a suborbital spacecraft~\cite{vessot}, and measurements using the latest cold atom clocks that verify gravity's effect on time with exquisite precision~\cite{2012PhRvL.109h0801G,2013PhRvA..87a0102P,2013PhRvL.111f0801L}.

\item
Tests of the equivalence principle for massive self-gravitating bodies (``Nordtvedt effect'') using lunar laser ranging~\cite{williams04}.

\item
Measurement of the ``dragging of inertial frames'' by the rotating Earth using gyroscopes in an orbiting satellite (Gravity Probe B)~\cite{2011PhRvL.106v1101E} and using laser tracking of Earth-orbiting satellites (LAGEOS)~\cite{2011EPJP..126...72C}.

\item
Tests of the constancy of Newton's gravitational constant, reaching a limit of one part in $10^{13}$ years using data from recent Mars orbiters~\cite{2011Icar..211..401K}.

\item
Stringent bounds on anomalous gravitational effects that are absent in GR, but present in alternative theories with preferred frames or intrusions of external gravity into local dynamics, using methods ranging from Earth-bound gravimeters to pulsar timing data~\cite{2012CQGra..29u5018S,2013CQGra..30p5019S}.

\item
Tests of the existence of gravitational radiation and tests of strong-field effects using binary pulsars (see the article by T. Damour in this volume).

\item
The total demise of Whitehead's theory, caused by its failure to pass five independent modern experimental tests~\cite{Gibbons200841}.
 
\end{itemize}

For an up-to-date review and references, see~\cite{2014LRR....17....4W}.  It is clear that experimental gravity has come a long way from the struggles to test Einstein's predictions in the years immediately following 1919.

\section{The gravitational lens: Einstein's gift to astronomy}
\label{sec:lenses}

In 1979, astronomers Dennis Walsh,
Robert Carswell and Ray Weymann discovered the ``double quasar''
Q0957+561, which consisted of two quasar images about 6 arcseconds
apart, with almost the same redshift ($z= 1.41$) and very similar
spectra~\cite{1979Natur.279..381W}.  Given that quasars are thought to be among the most distant
objects in the universe, the probability of finding two so close
together was low. It was immediately realized that there was just one
quasar, but that intervening matter in the form of a galaxy or a
cluster of galaxies was bending the light from the quasar and
producing two separate images.    
Since then, over 60 lensed quasars have been discovered.  

Ironically, Einstein was probably the first to consider the possibility of a gravitational lens, although he never published it.  Indeed, the fact that he did this calculation was unearthed only in 1997.  In the course of studying Einstein's original notebooks for the Einstein Papers Project, J\"urgen Renn and colleagues came across a notebook from around 1912, in which Einstein worked out the basic equations for gravitational lenses, including the possibility of double images for a point lens and the magnification of the intensity of the images~\cite{1997Sci...275..184R}.  Everything he did was off by a factor of two, of course, because he was using the pre-GR value for the deflection of light.  He concluded that the effects were too small and the probability of an astronomical lens too low ever to be of interest, so he never published the calculations.  In his 1920 book on general relativity, Eddington discussed the possibility of gravitational lenses~\cite{1920stga.book.....E}, and in 1924, Chwolson pointed out that perfect alignment between source, point lens and observer would lead to what is today called an ``Einstein ring'' image~\cite{1924AN....221..329C}.  In 1936, Einstein finally published a short note about gravitational lenses (using the correct numerical factor)~\cite{1936Sci....84..506E}, primarily, it seems, to get a Czech electrical engineer named Rudi Mandel to stop pestering him about it.  Fritz Zwicky pointed out that galaxies could act as gravitational lenses~\cite{1937PhRv...51..290Z}.   During the 1960s and 1970s, Jeno and Madeleine Barnothy argued persistently that lensing by foreground galaxies was responsible for the high luminosity of quasars~\cite{1968Sci...162..348B}, but this idea proved not to be valid.   

But since the discovery of the first gravitational lens, the phenomenon has been exploited
to map the distribution of mass around galaxies and clusters, and
to search for dark matter, dark energy, compact objects, and
extrasolar planets. Many subtopics of gravitational lensing have been
developed to cover different astronomical realms: microlensing for the
search of dim compact objects and extra-solar planets, the use of
luminous arcs to map the distribution of mass and dark matter, and
weak lensing to measure the properties of dark energy.  Lensing has to
be taken into account in interpreting certain aspects of the cosmic
microwave background radiation, and in extracting information from
gravitational waves emitted by sources at cosmological
distances. 

\begin{figure}
\centering
\includegraphics[width=4.5in]{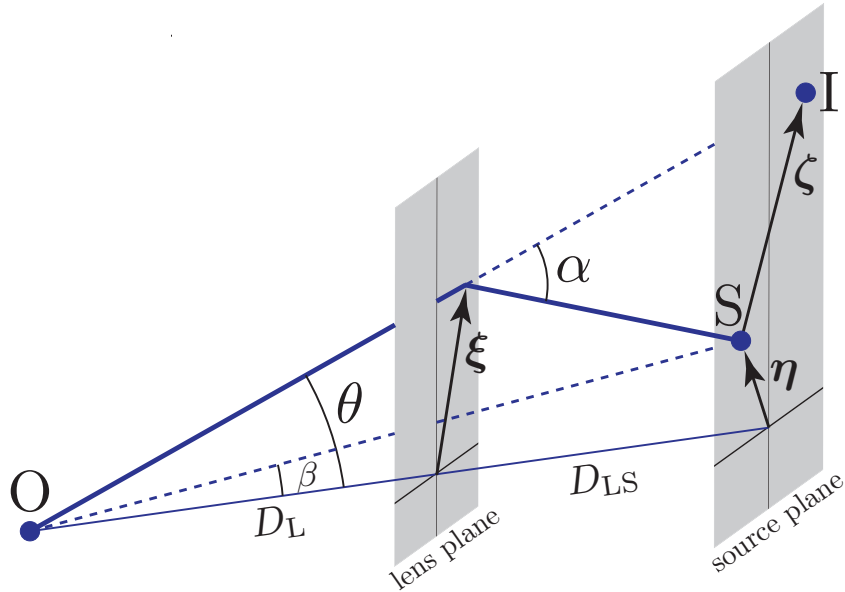}
\caption{Geometry of a gravitational lens. The observer is labeled O,
  the source S, and the image I.}
\label{fig:grav_lens}
\end{figure}

The basics of gravitational lensing may be established by referring to Fig.\ \ref{fig:grav_lens}.  A light ray emitted from the source S at a distance $D_S=D_L + D_{LS}$ is deflected by an angle $\alpha$ and received at O, making an angle $\theta$ relative to the selected mass element  of the lens.
Simple vector addition on the plane of the source yields $D_S {\bm \theta}  = {\bm \eta} + {\bm \zeta}$, with  ${\bm \eta} = D_S {\bm \beta}$ and ${\bm \zeta} = -D_{\rm LS}  {\bm \alpha}$, where $\bm \alpha$ is the deflection vector, given by
\begin{equation} 
\bm{\alpha}(\bm{\xi}) = -\frac{4G}{c^2}\int \Sigma(\bm{\xi'})  
\frac{\bm{\xi} - \bm{\xi'}}{|\bm{\xi} - \bm{\xi'}|^2}\, d^2\xi' \,, 
\label{eq11:alpha_lens} 
\end{equation} 
where $\Sigma(\bm{\xi'})$ is the projection of the mass density of the lens onto the lens plane.  The result is the lens equation
\begin{equation} 
\bm{\theta} + \frac{D_{\rm LS}}{D_{\rm S}} \bm{\alpha} = \bm{\beta}. 
\label{eq11:lens_eq} 
\end{equation} 
The simplest lens is that of a spherically symmetric body of mass $M$.  In this case the deflection vector is given by
\begin{equation} 
\bm{\alpha}(\bm{\xi}) =- \frac{4GM}{c^2} \frac{\bm{\xi}}{\xi^2} \,, 
\end{equation} 
and substitution into Eq.~(\ref{eq11:lens_eq}), along with $\xi = D_L \theta$, gives rise to a scalar equation 
$\theta - \theta_{\rm E}^2/\theta = \beta$, leading to two images with
\begin{equation} 
\theta_\pm = \frac{1}{2} \biggl( \beta 
\pm  \sqrt{ \beta^2 + 4 \theta_{\rm E}^2 } \biggr) \,,
\end{equation} 
where 
\begin{equation} 
\theta_{\rm E}^2 := \frac{4GM}{c^2} 
\frac{D_{\rm LS}}{D_{\rm S} D_{\rm L}} \,.
\label{eq11:Einstein_angle} 
\end{equation} 
The
parameter $\theta_{\rm E}$ is known as the {\it Einstein angle}, and  
the corresponding length scale 
\begin{equation} 
\xi_{\rm E} := D_{\rm L} \theta_{\rm E} 
= \sqrt{ \frac{4GM}{c^2} \frac{D_{\rm LS} D_{\rm L}}{D_{\rm S}} } 
\label{eq11:Einstein_radius} 
\end{equation} 
is the {\it Einstein radius}. For lenses of galactic scales
with $M \sim 10^{12} M_\odot$, $\theta_{\rm E} \simeq 1.8 \, \rm{as}$,
while for solar-mass lenses within the galaxy, $\theta_{\rm E} \simeq 0.5 \, \rm{mas}$.
For more complex lens mass distributions there can be multiple (typically odd numbers of) images.  For a thorough review of the mathematics of lensing and many of its applications, see~\cite{1992grle.book.....S}.

When the source has a nonzero angular size, the lens continues to 
displace its images, but there is also a distortion of its
shape. Points on opposite sides of the source perpendicular to the
optical axis are displaced by an angle $\theta$, and are therefore
stretched by a corresponding factor, while points on either side
parallel to the optical axis are stretched only by the 
difference in $\theta$.  A circular source can be distorted into an ellipse, or even into an arc with a convex side. 
The orientation and shapes of such luminous arcs have been used to
deduce the mass distribution of the galaxies or clusters that act as
lenses, a procedure sometimes dubbed {\em gravitational tomography} (see~\cite{1992ARA&A..30..311B} for a review). 
Even when the lens produces elliptical distortions that are too small
to be measured individually, there is a systematic effect, averaged
over large collections of images, that is sensitive to the evolution
of the universe over an epoch when dark energy began to be important;
this is the realm of weak gravitational lensing.  Weak lensing now plays a role in studies of large-scale structure, of quasar-galaxy correlations, and of fluctuations in the cosmic background radiation, and will be a key tool for probing dark energy (for reviews of weak lensing and its applications see~\cite{2001PhR...340..291B}).

Each image is altered in apparent brightness by a factor   
\begin{equation} 
\mu_\pm  
= \frac{1}{4} \left( 
\frac{\beta}{\sqrt{\beta^2 + 4\theta_{\rm E}^2}} 
+ \frac{\sqrt{\beta^2 + 4\theta_{\rm E}^2}}{\beta} 
\pm 2 \right ) \,.  
\end{equation}
If the images are too close together to be
resolved individually by the observer, the total
magnification is given by     
\begin{equation}
\mu_+ + \mu_- = 
\frac{1}{2} \left ( \frac{\beta}{\sqrt{\beta^2 + 4\theta_{\rm E}^2}} 
+ \frac{\sqrt{\beta^2 + 4\theta_{\rm E}^2}}{\beta} \right ), 
\label{eq11:lensluminosity}
\end{equation}
which is always greater than unity. This is the realm of microlensing.
The technique of monitoring the
variable brightness of lensed images was used in a series of
experiments to search for ``massive compact halo objects'' (MACHOs) in
our galaxy.  If the galaxy contains a population of dark objects
(black holes, neutron stars, brown dwarf stars, or other exotic
objects) with masses comparable to $M_\odot$, then the brightness of a
star transiting behind such an object should behave in a way
consistent with Eq.~(\ref{eq11:lensluminosity}). This effect can be  
distinguished from the star's own variability, or from the absorption
of starlight by intervening matter, because these tend to depend on
wavelength, while the lensing is independent of wavelength.  Searches
for dark objects passing in front of the dense field of stars in the
Large Magellanic Cloud and in the galactic center were carried out
between 1993 and 2007, placing a stringent upper limit on the amount
of halo mass that could be made up of such objects~\cite{2000ApJ...542..281A,2007A&A...469..387T}. This strengthened
the conclusion that the vast majority of the halo mass must be made of
non-baryonic dark matter.   

In 2003, an extra-solar planetary system was discovered by
microlensing~\cite{2004ApJ...606L.155B}.  The combined lensing of a distant source by a
Jupiter-scale companion and its host star was measured and could be
deconvolved to determine the mass ratio and the approximate distance
between the planet and the star. Additional systems were discovered
subsequently, and gravitational lensing is proving to be a useful tool in
the search for exoplanets.   

The first gravitational lensing of supernovae was reported in 2014~\cite{2014ApJ...786....9P,2014MNRAS.440.2742N}; three Type 1a supernovae from 2011 and 2012 were found to have had their brightness magnified by the lensing action of foreground galactic clusters.  In fact, gravitational lensing will be a complicating factor in efforts to use standard candles at large redshift to improve the Hubble relation and to measure cosmic acceleration, whether they be supernovae detected electromagnetically or inspiralling compact binaries detected by gravitational waves.

Finally, gravitational lensing yielded a remarkable test of the deflection of light on
galactic scales~\cite{2006PhRvD..74f1501B}. It used data on gravitational lensing by 15 elliptical
galaxies, collected by the Sloan Digital Sky Survey. The Newtonian
potential $U$ of each lensing galaxy (including the contribution from
dark matter) was derived from
the observed velocity dispersion of stars in the galaxy. Comparing
the observed lensing with the lensing predicted by the models provided a
10 percent bound on $\gamma$, in agreement with general
relativity.   Although the accuracy was only comparable to that of Eddington's 1919 measurements, this test of Einstein's light deflection was obtained
on a galactic, rather than solar-system scale.  

\section{Concluding remarks}
\label{sec:conclusions}

In 1919 the primary goal of scientists was to determine if this new theory that {\em The New York Times} said put stars ``all askew in the heavens'' and was comprehensible ``by no more than twelve wise men'', was correct.  Because, as the {\em London Times} put it, ``the scientific concept of the fabric of the universe must be changed'' in the face of general relativity, this was both a scientific question and a philosophical and human question.  Today at the centenary of the theory, we have learned to live with curved space-time, and we accept general relativity as the gold standard for gravity on solar-system and many astrophysical scales.  The goal now, while perhaps less epochal, is to search for possible new physics ``beyond Einstein'' that might occur at cosmological scales, in strong-field regimes, or where signatures of quantum gravity might be present.  Even if some modified version of general relativity must be adopted ultimately to accommodate new observations, the theory that caused such a sensation in 1919 will very likely still be its foundation.

\ack
This work was supported in part by the National Science Foundation,
Grant No.\ PHY 13-06069.

\section*{References}

\bibliographystyle{iopart-num}
\bibliography{refsMilestone}

\providecommand{\newblock}{}
\begin{thebibliography}{10}
\expandafter\ifx\csname url\endcsname\relax
  \def\url#1{{\tt #1}}\fi
\expandafter\ifx\csname urlprefix\endcsname\relax\def\urlprefix{URL }\fi
\providecommand{\eprint}[2][]{\url{#2}}

\bibitem{pauljohnson}
Johnson P 1983 {\em {Modern Times: The World from the Twenties to the
  Eighties}\/} (New York: Harper and Row)

\bibitem{1920RSPTA.220..291D}
{Dyson} F~W, {Eddington} A~S and {Davidson} C 1920 {\em Phil. Trans. Roy. Soc.
  London A\/} {\bf 220} 291--333

\bibitem{Crelinsten}
Crelinsten J 2006 {\em Einstein's Jury: The Race to Test Relativity\/}
  (Princeton: Princeton University Press)

\bibitem{1911AnP...340..898E}
{Einstein} A 1911 {\em Annalen der Physik\/} {\bf 340} 898--908

\bibitem{willcavendish}
Will C~M 1988 {\em Am. J. Phys.\/} {\bf 56} 413--415

\bibitem{1979Obs....99..195H}
{Harvey} G~M 1979 {\em The Observatory\/} {\bf 99} 195--198

\bibitem{2009PhT....62c..37K}
{Kennefick} D 2009 {\em Phys. Today\/} {\bf 62} 37

\bibitem{1960VA......3...47V}
{von Kl{\"u}ber} H 1960 {\em Vistas in Astronomy\/} {\bf 3} 47--77

\bibitem{BertottiBrillKrotkov}
Bertotti B, Brill D and Krotkov R 1962 Experiments on gravitation {\em
  Gravitation: An Introduction to Current Research\/} ed Witten L (New York:
  Wiley) pp 1--48

\bibitem{1976AJ.....81..452B}
{Brune} Jr R~A, {Cobb} C~L, {Dewitt} B~S, {Dewitt-Morette} C, {Evans} D~S,
  {Floyd} J~E, {Jones} B~F, {Lazenby} R~V, {Marin} M, {Matzner} R~A, {Mikesell}
  A~H, {Mikesell} M~R, {Mitchell} R~I, {Ryan} M~P, {Smith} H~J, {Sy} A and
  {Thompson} C~D 1976 {\em \aj\/} {\bf 81} 452--454

\bibitem{1976AJ.....81..455J}
{Jones} B~F 1976 {\em \aj\/} {\bf 81} 455--463

\bibitem{whitehead22}
Whitehead A~N 1922 {\em The Principle of Relativity, with Applications to
  Physical Science\/} (Cambridge: Cambridge University Press)

\bibitem{lopresto91}
LoPresto J~C, Schrader C and Pierce A~K 1991 {\em Astrophys. J.\/} {\bf 376}
  757--760

\bibitem{1960PhRvL...4..337P}
{Pound} R~V and {Rebka} G~A 1960 {\em \prl\/} {\bf 4} 337--341

\bibitem{lebach}
Lebach D~E, Corey B~E, Shapiro I~I, Ratner M~I, Webber J~C, Rogers A~E~E, Davis
  J~L and Herring T~A 1995 {\em Phys. Rev. Lett.\/} {\bf 75} 1439--1442

\bibitem{2009ApJ...699.1395F}
{Fomalont} E, {Kopeikin} S, {Lanyi} G and {Benson} J 2009 {\em \apj\/} {\bf
  699} 1395--1402 (\textit{Preprint} \eprint{0904.3992})

\bibitem{sshapiro04}
Shapiro S~S, Davis J~L, Lebach D~E and Gregory J~S 2004 {\em Phys. Rev.
  Lett.\/} {\bf 92} 121101

\bibitem{2009A&A...499..331L}
{Lambert} S~B and {Le Poncin-Lafitte} C 2009 {\em \aap\/} {\bf 499} 331--335
  (\textit{Preprint} \eprint{0903.1615})

\bibitem{2011A&A...529A..70L}
{Lambert} S~B and {Le Poncin-Lafitte} C 2011 {\em \aap\/} {\bf 529} A70

\bibitem{hipparcos}
Froeschl{\'{e}} M, Mignard F and Arenou F 1997 Determination of the ppn
  parameter $\gamma$ with the hipparcos data {\em Proceedings of the Hipparcos
  Venice Symposium\/} (Noordwijk, Netherlands: ESA)
  \urlprefix\url{http://astro.estec.esa.nl/Hipparcos/venice.html}

\bibitem{gaia}
Gaia - taking the galactic census
  \urlprefix\url{http://www.cosmos.esa.int/web/gaia/}

\bibitem{2002EAS.....2..107M}
{Mignard} F 2002 {Fundamental Physics with GAIA} {\em EAS Publications
  Series\/} vol~2 ed {Bienayme} O and {Turon} C pp 107--121

\bibitem{1964PhRvL..13..789S}
{Shapiro} I~I 1964 {\em \prl\/} {\bf 13} 789--791

\bibitem{reasenberg}
Reasenberg R~D, Shapiro I~I, MacNeil P~E, Goldstein R~B, Breidenthal J~C,
  Brenkle J~P, Cain D~L, Kaufman T~M, Komarek T~A and Zygielbaum A~I 1979 {\em
  Astrophys. J. Lett.\/} {\bf 234} L219--L221

\bibitem{bertotti03}
Bertotti B, Iess L and Tortora P 2003 {\em Nature\/} {\bf 425} 374--376

\bibitem{shapiro}
Shapiro I~I 1999 {\em Rev. Mod. Phys.\/} {\bf 71} S41--S53

\bibitem{2008ARNPS..58..207T}
{Turyshev} S~G 2008 {\em Ann. Rev. Nucl. Part. Sci.\/} {\bf 58} 207--248
  (\textit{Preprint} \eprint{0806.1731})

\bibitem{2010AmJPh..78.1240W}
{Will} C~M 2010 {\em Am. J. Phys.\/} {\bf 78} 1240--1247 (\textit{Preprint}
  \eprint{1008.0296})

\bibitem{2014LRR....17....4W}
{Will} C~M 2014 {\em Living Rev. Relativ.\/} {\bf 17} 4 (\textit{Preprint}
  \eprint{1403.7377})

\bibitem{Agashe:2014kda}
Olive K {\em et~al.\/} (Particle Data Group) 2014 {\em Chin.Phys.\/} {\bf C38}
  090001

\bibitem{2014A&A...561A.115V}
{Verma} A~K, {Fienga} A, {Laskar} J, {Manche} H and {Gastineau} M 2014 {\em
  \aap\/} {\bf 561} A115 (\textit{Preprint} \eprint{1306.5569})

\bibitem{vessot}
Vessot R, Levine M, Mattison E, Blomberg E, Hoffman T, Nystrom G, Farrell B,
  Decher R, Eby P, Baugher C, Watts J, Teuber D and Wills F 1980 {\em Phys.
  Rev. Lett.\/} {\bf 45} 2081--2084

\bibitem{2012PhRvL.109h0801G}
{Gu{\'e}na} J, {Abgrall} M, {Rovera} D, {Rosenbusch} P, {Tobar} M~E, {Laurent}
  P, {Clairon} A and {Bize} S 2012 {\em Phys. Rev. Lett.\/} {\bf 109} 080801
  (\textit{Preprint} \eprint{1205.4235})

\bibitem{2013PhRvA..87a0102P}
{Peil} S, {Crane} S, {Hanssen} J~L, {Swanson} T~B and {Ekstrom} C~R 2013 {\em
  Phys. Rev. A\/} {\bf 87} 010102 (\textit{Preprint} \eprint{1301.6145})

\bibitem{2013PhRvL.111f0801L}
{Leefer} N, {Weber} C~T~M, {Cing{\"o}z} A, {Torgerson} J~R and {Budker} D 2013
  {\em Phys. Rev. Lett.\/} {\bf 111} 060801 (\textit{Preprint}
  \eprint{1304.6940})

\bibitem{williams04}
Williams J, Turyshev S and Boggs D 2004 {\em Phys. Rev. Lett.\/} {\bf 93}
  261101 (\textit{Preprint} \eprint{gr-qc/0411113})

\bibitem{2011PhRvL.106v1101E}
{Everitt} C~W~F, {Debra} D~B, {Parkinson} B~W, {Turneaure} J~P, {Conklin} J~W,
  {Heifetz} M~I, {Keiser} G~M, {Silbergleit} A~S, {Holmes} T, {Kolodziejczak}
  J, {Al-Meshari} M, {Mester} J~C, {Muhlfelder} B, {Solomonik} V~G, {Stahl} K,
  {Worden} Jr P~W, {Bencze} W, {Buchman} S, {Clarke} B, {Al-Jadaan} A,
  {Al-Jibreen} H, {Li} J, {Lipa} J~A, {Lockhart} J~M, {Al-Suwaidan} B, {Taber}
  M and {Wang} S 2011 {\em Phys. Rev. Lett.\/} {\bf 106} 221101
  (\textit{Preprint} \eprint{1105.3456})

\bibitem{2011EPJP..126...72C}
{Ciufolini} I, {Paolozzi} A, {Pavlis} E~C, {Ries} J, {Koenig} R, {Matzner} R,
  {Sindoni} G and {Neumeyer} H 2011 {\em Eur. Phys. J. Plus\/} {\bf 126} 72

\bibitem{2011Icar..211..401K}
{Konopliv} A~S, {Asmar} S~W, {Folkner} W~M, {Karatekin} {\"O}, {Nunes} D~C,
  {Smrekar} S~E, {Yoder} C~F and {Zuber} M~T 2011 {\em Icarus\/} {\bf 211}
  401--428

\bibitem{2012CQGra..29u5018S}
{Shao} L and {Wex} N 2012 {\em Class. Quantum Grav.\/} {\bf 29} 215018
  (\textit{Preprint} \eprint{1209.4503})

\bibitem{2013CQGra..30p5019S}
{Shao} L, {Caballero} R~N, {Kramer} M, {Wex} N, {Champion} D~J and {Jessner} A
  2013 {\em Class. Quantum Grav.\/} {\bf 30} 165019 (\textit{Preprint}
  \eprint{1307.2552})

\bibitem{Gibbons200841}
Gibbons G and Will C~M 2008 {\em Studies in History and Philosophy of Modern
  Physics\/} {\bf 39} 41 -- 61 (\textit{Preprint} \eprint{gr-qc/0611006})

\bibitem{1979Natur.279..381W}
{Walsh} D, {Carswell} R~F and {Weymann} R~J 1979 {\em \nat\/} {\bf 279}
  381--384

\bibitem{1997Sci...275..184R}
{Renn} J, {Sauer} T and {Stachel} J 1997 {\em Science\/} {\bf 275} 184--186

\bibitem{1920stga.book.....E}
{Eddington} A~S 1920 {\em {Space, Time and Gravitation. an Outline of the
  General Relativity Theory}\/} (Cambridge University Press)

\bibitem{1924AN....221..329C}
{Chwolson} O 1924 {\em Astron. Nachr.\/} {\bf 221} 329

\bibitem{1936Sci....84..506E}
{Einstein} A 1936 {\em Science\/} {\bf 84} 506--507

\bibitem{1937PhRv...51..290Z}
{Zwicky} F 1937 {\em Phys. Rev.\/} {\bf 51} 290--290

\bibitem{1968Sci...162..348B}
{Barnothy} J and {Barnothy} M~F 1968 {\em Science\/} {\bf 162} 348--352

\bibitem{1992grle.book.....S}
{Schneider} P, {Ehlers} J and {Falco} E~E 1992 {\em {Gravitational Lenses}\/}
  (Berlin: Springer-Verlag)

\bibitem{1992ARA&A..30..311B}
{Blandford} R~D and {Narayan} R 1992 {\em \araa\/} {\bf 30} 311--358

\bibitem{2001PhR...340..291B}
{Bartelmann} M and {Schneider} P 2001 {\em \physrep\/} {\bf 340} 291--472
  (\textit{Preprint} \eprint{astro-ph/9912508})

\bibitem{2000ApJ...542..281A}
{Alcock} C, {Allsman} R~A, {Alves} D~R, {Axelrod} T~S, {Becker} A~C, {Bennett}
  D~P, {Cook} K~H, {Dalal} N, {Drake} A~J, {Freeman} K~C, {Geha} M, {Griest} K,
  {Lehner} M~J, {Marshall} S~L, {Minniti} D, {Nelson} C~A, {Peterson} B~A,
  {Popowski} P, {Pratt} M~R, {Quinn} P~J, {Stubbs} C~W, {Sutherland} W,
  {Tomaney} A~B, {Vandehei} T and {Welch} D 2000 {\em \apj\/} {\bf 542}
  281--307 (\textit{Preprint} \eprint{astro-ph/0001272})

\bibitem{2007A&A...469..387T}
{Tisserand} P, {Le Guillou} L, {Afonso} C, {Albert} J~N, {Andersen} J, {Ansari}
  R, {Aubourg} {\'E}, {Bareyre} P, {Beaulieu} J~P, {Charlot} X, {Coutures} C,
  {Ferlet} R, {Fouqu{\'e}} P, {Glicenstein} J~F, {Goldman} B, {Gould} A,
  {Graff} D, {Gros} M, {Haissinski} J, {Hamadache} C, {de Kat} J, {Lasserre} T,
  {Lesquoy} {\'E}, {Loup} C, {Magneville} C, {Marquette} J~B, {Maurice} {\'E},
  {Maury} A, {Milsztajn} A, {Moniez} M, {Palanque-Delabrouille} N, {Perdereau}
  O, {Rahal} Y~R, {Rich} J, {Spiro} M, {Vidal-Madjar} A, {Vigroux} L,
  {Zylberajch} S and {EROS-2 Collaboration} 2007 {\em \aap\/} {\bf 469}
  387--404 (\textit{Preprint} \eprint{astro-ph/0607207})

\bibitem{2004ApJ...606L.155B}
{Bond} I~A, {Udalski} A, {Jaroszy{\'n}ski} M, {Rattenbury} N~J, {Paczy{\'n}ski}
  B, {Soszy{\'n}ski} I, {Wyrzykowski} L, {Szyma{\'n}ski} M~K, {Kubiak} M,
  {Szewczyk} O, {{\.Z}ebru{\'n}} K, {Pietrzy{\'n}ski} G, {Abe} F, {Bennett}
  D~P, {Eguchi} S, {Furuta} Y, {Hearnshaw} J~B, {Kamiya} K, {Kilmartin} P~M,
  {Kurata} Y, {Masuda} K, {Matsubara} Y, {Muraki} Y, {Noda} S, {Okajima} K,
  {Sako} T, {Sekiguchi} T, {Sullivan} D~J, {Sumi} T, {Tristram} P~J,
  {Yanagisawa} T, {Yock} P~C~M and {OGLE Collaboration} 2004 {\em \apjl\/} {\bf
  606} L155--L158 (\textit{Preprint} \eprint{astro-ph/0404309})

\bibitem{2014ApJ...786....9P}
{Patel} B, {McCully} C, {Jha} S~W, {Rodney} S~A, {Jones} D~O, {Graur} O,
  {Merten} J, {Zitrin} A, {Riess} A~G, {Matheson} T, {Sako} M, {Holoien} T~W~S,
  {Postman} M, {Coe} D, {Bartelmann} M, {Balestra} I, {Ben{\'{\i}}tez} N,
  {Bouwens} R, {Bradley} L, {Broadhurst} T, {Cenko} S~B, {Donahue} M,
  {Filippenko} A~V, {Ford} H, {Garnavich} P, {Grillo} C, {Infante} L, {Jouvel}
  S, {Kelson} D, {Koekemoer} A, {Lahav} O, {Lemze} D, {Maoz} D, {Medezinski} E,
  {Melchior} P, {Meneghetti} M, {Molino} A, {Moustakas} J, {Moustakas} L~A,
  {Nonino} M, {Rosati} P, {Seitz} S, {Strolger} L~G, {Umetsu} K and {Zheng} W
  2014 {\em \apj\/} {\bf 786} 9 (\textit{Preprint} \eprint{1312.0943})

\bibitem{2014MNRAS.440.2742N}
{Nordin} J, {Rubin} D, {Richard} J, {Rykoff} E, {Aldering} G, {Amanullah} R,
  {Atek} H, {Barbary} K, {Deustua} S, {Fakhouri} H~K, {Fruchter} A~S, {Goobar}
  A, {Hook} I, {Hsiao} E~Y, {Huang} X, {Kneib} J~P, {Lidman} C, {Meyers} J,
  {Perlmutter} S, {Saunders} C, {Spadafora} A~L, {Suzuki} N and {Supernova
  Cosmology Project} 2014 {\em \mnras\/} {\bf 440} 2742--2754
  (\textit{Preprint} \eprint{1312.2576})

\bibitem{2006PhRvD..74f1501B}
{Bolton} A~S, {Rappaport} S and {Burles} S 2006 {\em \prd\/} {\bf 74} 061501
  (\textit{Preprint} \eprint{astro-ph/0607657})

\end{thebibliography}

\end{document}